\documentclass[epj]{svjour}
\usepackage{graphicx}
\begin{document}
\title{Phase transitions in quantum tunneling and quasi-exact solvability for a family of parametric double-well potentials}
\titlerunning{Quantum tuneling in deformable DW potentials}
\author{F. Naha Nzoupe\inst{1,2} \and Alain M. Dikand\'e\inst{2} \and C. Tchawoua\inst{1}
}                     
\offprints{dikande.alain@ubuea.cm}          
\institute{Laboratory of Mechanics, Department of
Physics, Faculty of Science, University of Yaound\'e I
P.O. Box 812 Yaound\'e, Cameroon \and Laboratory of Research on Advanced Materials and Nonlinear Sciences (LaRAMaNS),
Department of Physics, Faculty of Science, University of Buea P.O. Box 63 Buea, Cameroon}
\date{Received: date / Revised version: date}
%
\abstract{
A parametrized double-well potential is proposed to address the issue of the impact of shape deformability of some bistable physical systems, on their quantum dynamics and classical statistical mechanics. The parametrized double-well potential possesses two fixed degenerate minima and a constant barrier height, but a tunable shape of its walls influencing the confinement of the two symmetric wells. The transition from quantum tunneling to classical crossover is investigated and it is found that unlike the bistable model based on the Ginzburg-Landau energy functional, so-called $\phi^4$ potential model, members of the family of parametrized double-well potentials can display a first-order transition above a universal critical value of the shape deformability parameter. Addressing their statistical mechanics within the framework of the transfer-integral operator formalism, the classical partition function is constructed with emphasis on low-lying eigenstates of the resulting Schr\"odinger-type equation. The quasi-exact solvability of this last equation to find the exact expression of the partition function, is shown to depend on a specific relationship between the shape deformability parameter and the thermodynamic temperature. The quasi-exact solvability condition is expressed analytically, and with this condition some exact wavefunctions and corresponding eigen energies are derived at several temperatures. The exact probability densities obtained from the analytical expressions of the groundstate wavefunctions at different temperatures, are found to be in excellent agreement with the probability density from the Langevin dynamics. This agreement makes it possible to study probability density-based thermodynamics via Langevin techniques.
} 
\maketitle
\section{Introduction}
\label{intro}
Instantons \cite{1,2} are elementary excitations with soliton features \cite{3,4} that play an important role in tunneling processes in condensed matter at low temperatures. Indeed, in multistate systems, transitions between two states separated by a potential barrier arise either in the classical regime due to thermal activation,  or in the quantum regime due to tunnelings. At high temperatures the classical thermal activation governs transitions occuring as a hopping over a potential barrier, whereas at low enough temperatures (i.e. near quantum criticality $T\rightarrow 0$) the transition is governed by a quantum tunneling through the potential barrier. In this later case the system dynamics is governed by classical configurations called instantons, known to dominate the thermal rate at low temperatures. As temperature rises the thermally induced crossover becomes more and more important, and at some critical temperature a phenomenon identified as phase transition in quantum tunneling \cite{5} takes place. This phenomenon has long attracted a great deal of interest, in particular it has been demonstrated that physical systems can exhibit not only a smooth second-order transition at a critical temperature $T_0$, but also a first-order transition \cite{6,7,8,9,10,11,12}. It should be noted that the first-order transition is related to the variation of the instanton action with the instanton energy.\\ Another aspect of interest in the study of elementary excitations and solitary waves in general, has been their contributions to the low-temperature statistical mechanics of nonlinear systems \cite{krum1,13a,13,13b,13c}. In ref. \cite{13} a general ideal kink-gaz phenomenology was developed to investigate the low-temperature statistical mechanics of systems admitting kink solitary waves and solitons. This phenomenology, so-called transfer-integral approach and restricted initially to systems for which kink-phonon interactions involve transparent scattering processes, was extended in ref. \cite{14} to non-reflectionless kink bearing nonlinear Klein-Gordon models. \\
However, while the above mentioned studies involve mostly two universal models namely the sine-Gordon \cite{17,18} model assumed to describe systems with periodic one-site potentials, and the $\phi^4$ \cite{19,20} model intended for physical systems with double-well (DW) potentials, real physical systems to which the studies address are rather quite diverse and most often uniques in some aspects of their physical features. To this last point, the sine-Gordon and $\phi^4$ potentials have fixed (i.e. universal) extrema while their shape profiles are rigids, restricting their applications to very few physical systems. To lift the shortcomings related to the regidity of their shape profiles, these two universal models have been parametrized. As a matter of fact the sine-Gordon potential was generalized by Remoissenet and Peyrard \cite{3,21,22} into a parametrized periodic potential (well-known as Remoissenet-Peyrard potential) for which the extrema, period and shape profiles can be tuned at wish by varying a deformability parameter \cite{21,22}. Similarly the $\phi^4$ potential was generalized to meet the requirement of tunable extrema and shape deformability of the DW potential \cite{15,23,24,25}. Let us stress that although some other patrametric DW potentials are reported in the literature as for instance the double-Morse \cite{dm1,39,34}, the family of DW potentials proposed in \cite{15,23,24,25} is peculiar in that it groups three distinct classes with distinct shape deformability features, but the three classes reduce to the $\phi^4$ potential when the deformability parameter tends to zero.\\ 
The importance of shape deformability in the context of bistable systems, lies in two issues related to structural properties of bistable fields in general. The first has to do with symmetry breaking phenomena, for which the conventional $\phi^{4}$ model predicts transitions in quatum tunneling only of the second order \cite{9,10,11,12}. The second is related to findings \cite{krum1,13} that the transfer-integral formalism always reduces the classical statistical mechanics of a one-dimensional (1D) $\phi^4$-field theory, to a time-dependent quantum-mechanical problem for which exact solutions do not exist. The first issue was recently adressed by Zhou et al. \cite{26}, who formulated the problem of transitions in quantum tunneling for one of the three classes of the Dikand\'e-Kofan\'e DW potentials \cite{23}. They found that due to an extra degree of freedom introduced by shape parametrization of the bistable system, the parametrized DW system could exhibit a first-order transition occuring at a finite critical value of the shape deformability parameter, besides the second-order transition predicted within the framework of the standard  $\phi^4$ field. Concerning the second issue, motivated by the fact that the classical statistical mechanics of a field-theoretical system can be fully analyzed with just the knowledge of its low-lying eigenstates, parametrized DW models are quite likely to introduce the possibility for quasi-exactly solvable (QES) systems in field theory as established recently in some contexts \cite{27,28,29,30,31}. \\
In this work we wish to address the above two issues namely the possible occurence of first-order transition in quantum tunneling and the exact solvability of the transfer-integral operator problem in statistical mechanics, for a bistable model whose DW potential is parametrized in such a way that only the confinement of potential wells and steepness of the potential walls are affected, leaving unaffected the barrier height as well as positions of the two degenerate potential minima. We first investigate phase transitions and determine conditions for the occurence of a first-order transition in quantum tunneling, in terms of a critical value of the shape deformability parameter. Next the classical statistical mechanics of the model is constructed within the framework of the transfer-integral aoperator pproach,and from the resulting  Schr\"odinger-like equation we establish that quasi-exact solbability holds for the model through a specific requirement relating temperature with the shape deformability parameter. Analytical expressions of some low-lying eigenfunctions and the corresponding energies are given, and results for the probability density are discussed both analytically and via numerical simulations of the Langevin equation with Gaussian white noise.

\section{The parametrized DW potential}
\label{two}
\begin{figure} 
\includegraphics[width=3.5in,height=2.5in]{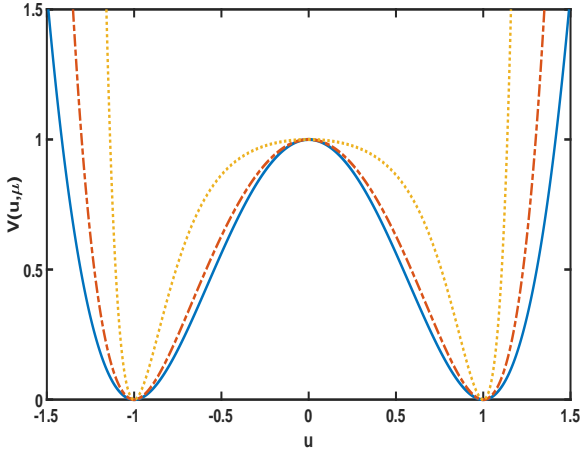}
\caption{(Color online) Profiles of the parametrized DW potential $V(u, \mu)$, for different values of the shape deformability parameter $\mu$:  $\mu\rightarrow 0$ (Solid line), $\mu=1.0$ (Dot-dashed line),  $\mu=4.0$ (Dotted line). $a_{0} = 1$.} 
\label{fig1}
\end{figure}

We are interested in the dynamics of $1D$ systems whose one-body potentials can be described by a class of parametrized DW potentials of the form:
\begin{equation}
V(u,\mu) = a(\mu)\left( \frac{\sinh^{2}(\alpha(\mu) u)}{\mu^{2}} - 1\right)^{2}, \hskip 0.3truecm  \mu > 0,  
\label{e1}
\end{equation}
where $a(\mu)>0$ and $\alpha(\mu)$ are two functions of the shape deformability parameter $\mu$. $V(u,\mu)$ belongs to a family of parametric DW potentials \cite{15,23,24,25} whose shape profiles can be tuned differently, but which admit a common asymptotic limit i.e. the $\phi^4$ potential. For the new member the two functions $a(\mu)$ and $\alpha(\mu)$ are defined as:
 \begin{equation}
\alpha(\mu) = sinh^{-1}\mu, \hskip 0.3truecm a(\mu) = a_0,
\label{e2}
\end{equation} 
in which case $V(u,\mu)$ is a DW potential with two degenerate minima fixed at $u = \pm 1$, and a barrier height also fixed at a constant value $a_0$. However a variation of $\mu$ changes the steepness of the potential walls, and consequently the sharpness (or confinement) of the potential wells. In fig. \ref{fig1} $V(u,\mu)$ is sketched for some arbitrary values of $\mu$. When $\mu$ tends to zero, the parametrized DW potential reduces to the universal bistable potential \cite{krum1,13} $V(u)=a_0\left(u^2 - 1\right)^2$. 
When $\mu$ is varied the minima positions and the barrier height remain unchanged, however the slope of the potential walls gets steeper. Hence the narrowest part of the potential barrier broadens while the flatness of the barrier top beocomes more pronounced, resulting in an enhancement of the confinement of the potential wells. \\
A variation of the barrier shape as observed in fig. \ref{fig1}, is consistent with the trend in both the experimentally observed rates and experimentally observed kinetic isotope effects, when studying vibrationally assisted hydrogen tunneling in enzyme-catalyzed reactions \cite{33}. Indeed, the effect of the catalyst becoming less efficient results in a progressive broadening of the narrowest part of the barrier, with the shoulder shape of the barrier peak becoming increasingly less pronounced. The patrametrized DW potential eqs. (\ref{e1})-(\ref{e2}) is infinite at the boundaries of the final interval and so is suitable for modeling hydrogen bonds in enzyme-catalyzed reactions. Also worthwhile to note, for significant values of the shape deformability parameter $\mu$, the DW potential eqs. (\ref{e1})-(\ref{e2}) displays features similar to the double-Morse potential used \cite{34} to describe proton motion in the hydrogen bond $O-H\cdot\cdot\cdot O$ in KDP ferroelectrics. In these specific materials \cite{34} the one-body proton potential rises steeply in the vicinity of the oxygen atoms, with a gentler slope at the sides of the potential barrier. The parametric DW potential eqs. (\ref{e1})-(\ref{e2}) can reproduce some of these peculiar features, namely by taking the proton displacement $u$ from the center of the hydrogen bond and using $\mu$ to mimic the rate of variation of the $O-O$ bond distance. Last but not least, a confinement of potential wells in bistable systems has recently been shown to hold a crucial role in the occurence of stochastic resonance in these systems \cite{35}. Therefore the parametric DW potential $V(u,\mu)$ is itself of great interest in Condensed Matter Physics. 

\section{First-order transition in quantum tunneling}
\label{three}
Consider a $1D$ quantum system with an Euclidean action (in dimensionless form):
\begin{equation}
S = \int d\tau\left(\frac{1}{2}\left(\frac{du}{d\tau}\right)^{2}+ V(u,\mu)\right),
\label{e3}
\end{equation}
 where $u$ is a scalar field in one time and zero space dimension, $\tau= i t$ is the imaginary time and $V(u,\mu)$ is the parametrized DW potential energy. The integral is taken over the period $\tau_{p}$ of the path. In statistical mechanics this period is related to temperature $T$ through the relation $\tau_{p} = \hbar /(k_{B}T)$, where $k_{B}$ is the Boltzmann constant. Without loss of generalities we will take $\hbar\equiv 1$. \\
 With an exponential accuracy the decay rate of the system in the semiclassical limit can be expressed:
\begin{equation}
\Gamma \sim \exp^{-F_{min}/T},
\label{e4}
\end{equation} 
where $F_{min}$ is the minimum of the effective "free energy" \cite{10} $F\equiv E + T S(E) - E_{min}$, with $E$ the energy of a classical pseudoparticle in the system while $E_{min}=0$ corresponds to the bottom of the potential. The minimum of the effective Euclidean action i.e. $S_{min}$, is obtained by minimizing (\ref{e3}) along the trajectories $\phi(\tau)$ (with period  $\tau_{p}$) satisfying the energy-integral equation:
\begin{equation}
\left(\frac{du_{c}}{dt}\right)^{2}= 2\left( V(u_{c},\mu )- E\right).
\label{e5}
\end{equation}
When $E=0$, corresponding to $\tau_{p}= \infty$ and $T=0$, the particle is at rest at the bottom of one of the two degenerate potential wells. The solution of (\ref{e5}) is a regular vacuum instanton (kink soliton) given by:
\begin{equation}
u_{c}(\tau)\rightarrow  \frac{1}{\alpha(\mu)}\tanh^{-1}\left[\frac{\mu}{\sqrt{1+\mu^{2}}} \tanh \frac{\tau}{\sqrt{2}d(\mu)}\right],
\label{e6}
\end{equation}
where $d(\mu)= \mu\left[a_{0}\alpha^{2}(\mu)(1+\mu^{2})\right]^{-1/2}$ is the kink width. Instead imposing periodic boundary conditions, with $\tau_{p}$ the period of motion, leads to the following expression for the trajectory $u_{c}(\tau)$ with energy $E\geq 0$:
\begin{equation}
u_{c}(\tau)=  \frac{1}{\alpha(\mu)}\tanh^{-1}\left[C_{1}\cdot sn\left(C_{2}\tau,\kappa \right)\right],
\label{e7}
\end{equation}
with $sn(\tau,\kappa )$ a Jacobi elliptic function \cite{jac1}, the modulus $\kappa$ of which is defined as: 
 \begin{equation}
\kappa = \sqrt{\frac{\left(1-\sqrt{E/a_{0}}\right)\left[1+\mu^{2}\left(1+\sqrt{E/a_{0}}\right)\right]}{\left(1+\sqrt{E/a_{0}}\right)\left[1+\mu^{2}\left(1-\sqrt{E/a_{0}}\right)\right]}}. 
\label{e8}
\end{equation}
The two parameters $C_{1}$ and $C_{2}$ appearing in formula (\ref{e7}) were defined as: 
\begin{equation}
C_{1} =\sqrt{\frac{\mu^{2}\left(1-\sqrt{E/a_{0}}\right)}{1+\mu^{2}\left(1-\sqrt{E/a_{0}}\right)}},
\label{e9}
\end{equation}
\begin{equation}
C_{2} = \frac{\alpha(\mu)}{\mu}\sqrt{2a_{0}\left(1+\sqrt{ \frac{E}{a_{0}} }\right)\left[1+\mu^{2}\left(1-\sqrt{  \frac{E}{a_{0}}}\right)\right]}.
\label{e10}
\end{equation}
The trajectory (\ref{e7}) possesses real periods for values of its argument equal to $4m\mathcal{K}(\kappa)$, where $m$ is an integer and $\mathcal{K}(\kappa)$ is the quarter period determined by the complete elliptic integral of the first kind \cite{jac1}. Eq. (\ref{e7}) therefore describes a periodic trajectory which we can refer to as periodon, the period of which is $(m = 1)$:
\begin{equation}
\tau_p = \frac{4}{C_2}\mathcal{K}(\kappa). \label{e11}
\end{equation}
The classical action corresponding to the periodon eq. (\ref{e7}) is obtained as:
\begin{equation}
S_{p}(E) = E\,\tau_{p} + W\left( u_{c}(\tau_{p})/2, E  \right),
\label{e12}
\end{equation}
where:
\begin{eqnarray}
W\left(u_c(\tau_{p})/2, E  \right)&=& \frac{2 C_{2}}{(\alpha(\mu) C_{1})^{2}}\lbrack (C_{1}^{4}-\kappa^{2})\Pi(C_{1},\kappa) \nonumber \\
                          &+& \kappa^{2}\mathcal{K}(\kappa)+ C_{1}^{2}(\mathcal{K}(\kappa) - \mathcal{E}(\kappa))\rbrack. \nonumber \\
                          \label{e13}
\end{eqnarray}
$\mathcal{E}(\kappa)$ and $\Pi(C_{1},\kappa)$ in the last formula are the complete elliptic integrals of the second and third kinds, respectively \cite{jac1}. \\
At $E = a_0$, which corresponds to the top of the potential barrier, the solution of eq. (\ref{e5}) is the trivial configuration $u_c(\tau)=0$. This trajectory is the sphaleron \cite{36} and its action is the thermodynamic action namely:
\begin{equation}
S_{0}(\mu) = a_{0}\tau.
\label{e14}
\end{equation}
 For the sphaleron the escape rate has the Boltzmann signature characteristic of a pure thermal activation, i.e.:
 \begin{equation}
\Gamma_{c} \sim \exp\left(- a(\mu)\tau_{p}\right) = \exp\left(- a(\mu)/k_B T\right).
\label{e15}
\end{equation}
Based on the above discussions, we can conclude that a periodon interpolates between the sphaleron $u_{c}(\tau)=0$ and the instanton (\ref{e6}). Hence the escape rate in the preriodon sector will be: 
\begin{equation}
\Gamma \sim \exp\left(- S_{min}(E)\right),
\label{e16}
\end{equation}
where $S_{min}(E) = min\left\lbrace S_{0}, S_{p}(E) \right\rbrace$. \\
To discuss the possible occurence and characteristic features of the transition(s) in quantum tunneling for the $1D$ quantum system with the action given by (\ref{e3}), it is useful to start by remarking that for periodic problems in classical statistical mechanics, the derivative of the action with respect to the energy is equivalent to the oscillation time $\tau$ of the system with this energy. For the oscillation time $\tau$ is proportional to the inverse temperature, with the action corresponding to motion in the periodon sector one can readily define the period of motion as:
\begin{equation}
\tau_{p}(E) = \frac{1}{k_BT}= \frac{dS_{p}(E)}{dE}, \hskip 0.3truecm  \frac{dS_{0}}{d\tau_{p}} = a(\mu).
\label{e17}
\end{equation}
Taking (\ref{e17}) together with (\ref{e11}) and (\ref{e12}), we can analyze the impact of the shape deformability parameter $\mu$ on the temperature dependence of $S_{min}$, based upon the energy dependence of the period of periodon $\tau_{p}(E)$. In particular these equations should permit us obtain the critical value of $\mu$ at which a first-order transition from quantum to thermal regime could be expected.\\
Two well-established criteria are known \cite{5,8} that determine conditions for occurence of transitions in quantum tunnelings, in the general problem of decay of a metastable state \cite{5,8,9,10,26a}: the first states that the transition will be of first order if the period $\tau_{p}(E)$ decreases to a minimum, and then rises again when $E$ increases from the potential bottom to the barrier height. If $\tau_{p}(E)$ instead decreases monotonically with increasing $E$ the transition is of second order. The second criteria states that if at some critical temperature the first derivative of $S_{min}(T)$ is discontinuous and an abrupt change is observed in the temperature dependence of the action, then the transition from quantum to thermal regime is a first-order transition in temperature. \\
The first criteria is equivalent to solving the equation:
\begin{equation}
\frac{d}{dE}\tau_{p}(E)=0, \label{17a}
\end{equation}
seeking for nontrivial solutions, with the temperature dependence of $\tau_{p}(E)$ given by (\ref{e17}). We solved the equation numerically by setting $a_{0}= 1/2$, and evaluated the energy $E_1$ corresponding to the minimum of $\tau_{p}(E)$ for some values of $\mu$. Results are shown in table \ref{table1}.

\begin{figure} 
\includegraphics[width=3.5in,height=2.5in]{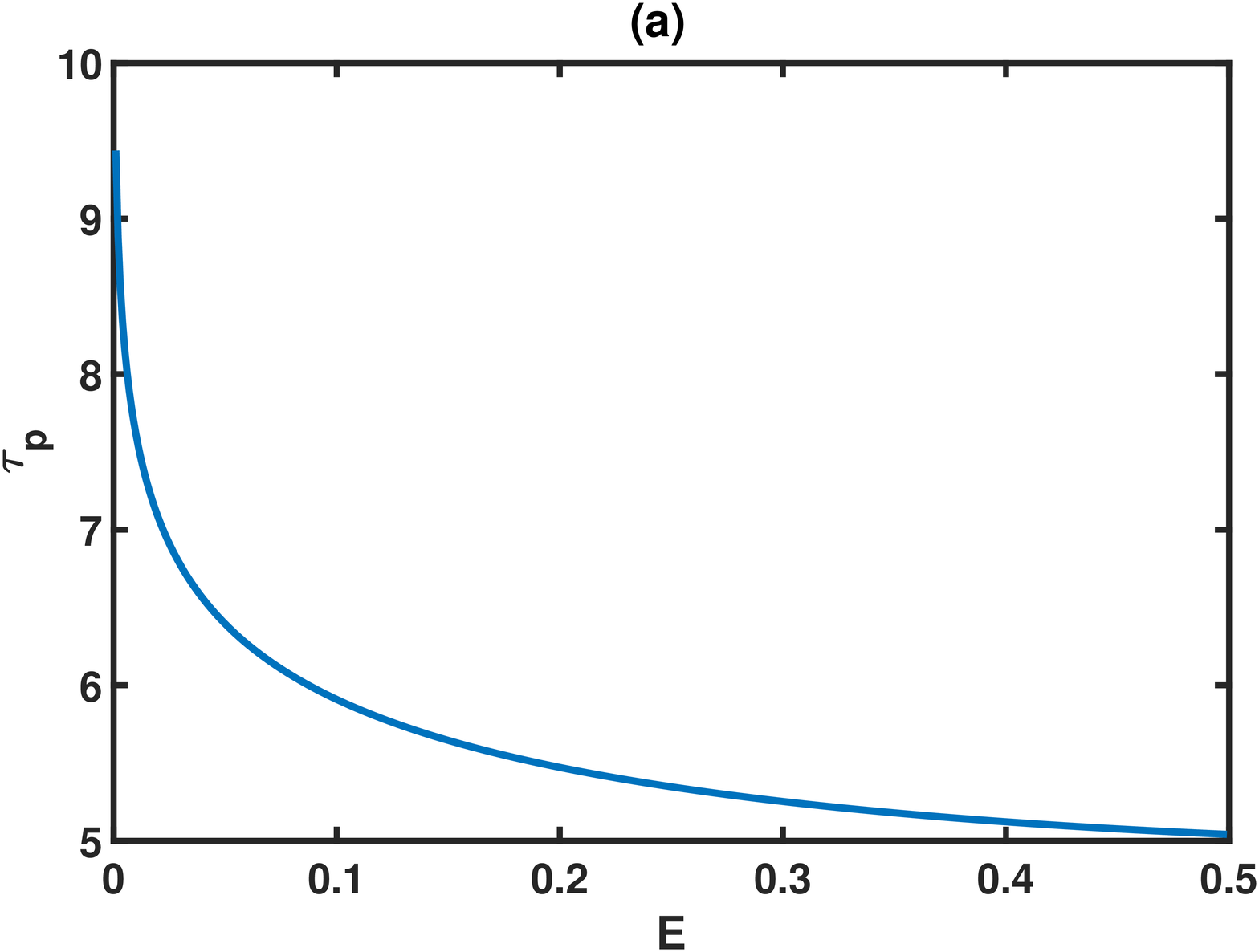}
\includegraphics[width=3.5in,height=2.5in]{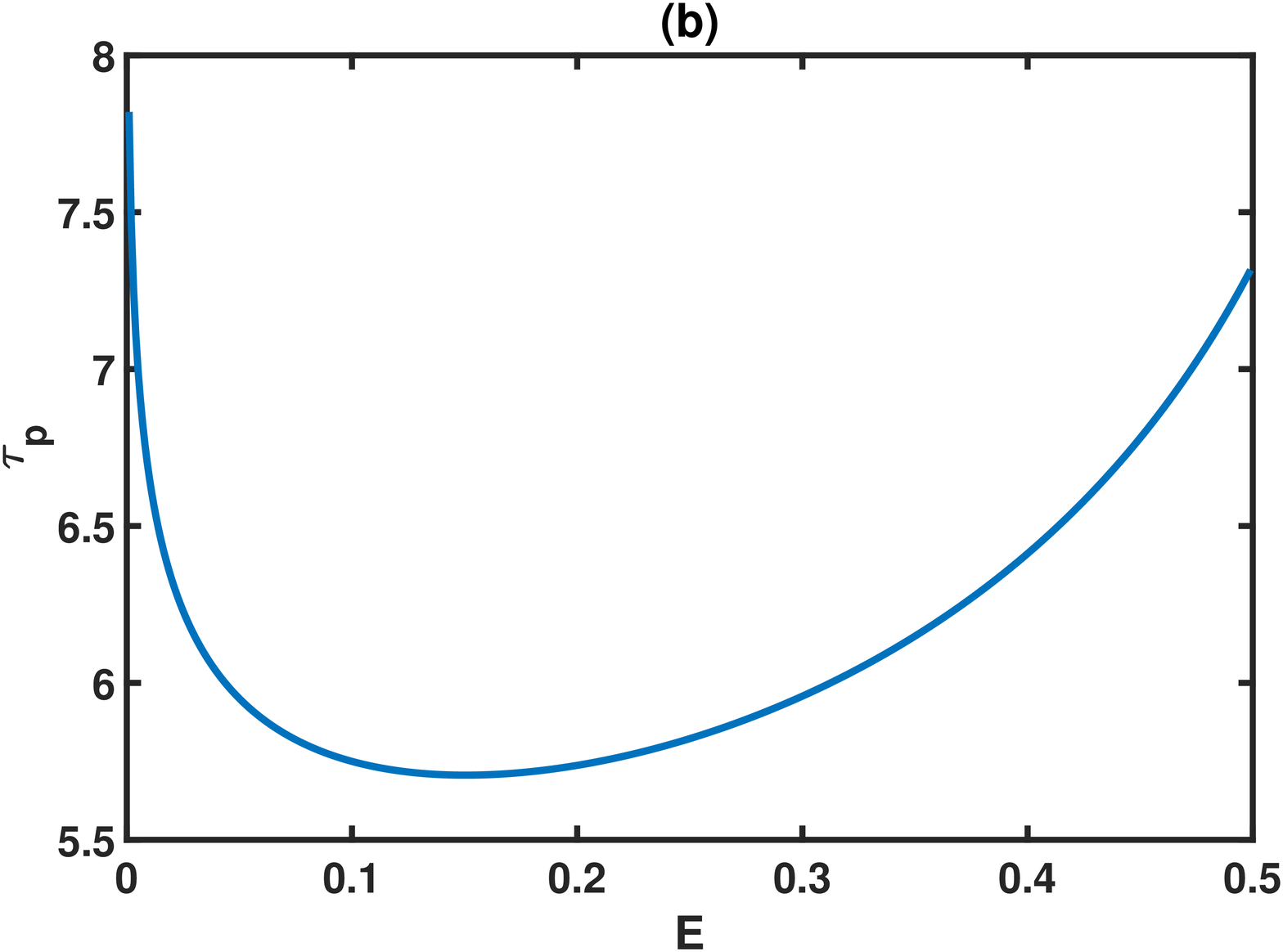}
\caption{(Color online) Variation of the instanton period with the energy $E$ for (a) $\mu = 1$, and (b) $\mu = 3$. Here $a_0=0.5$. } 
\label{fig2}
\end{figure}

\begin{figure} 
\includegraphics[width=3.5in,height=2.5in]{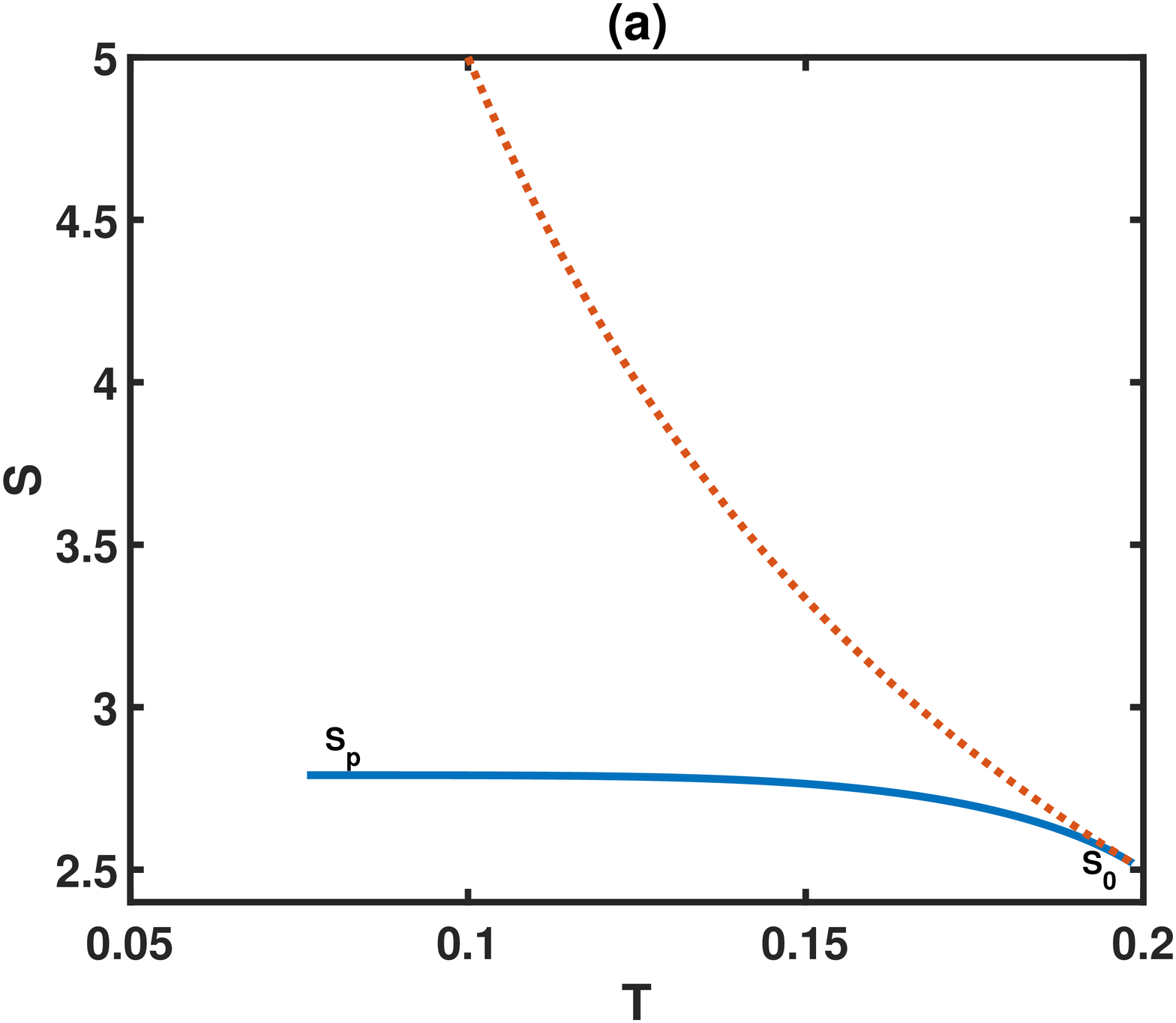}
\includegraphics[width=3.5in,height=2.5in]{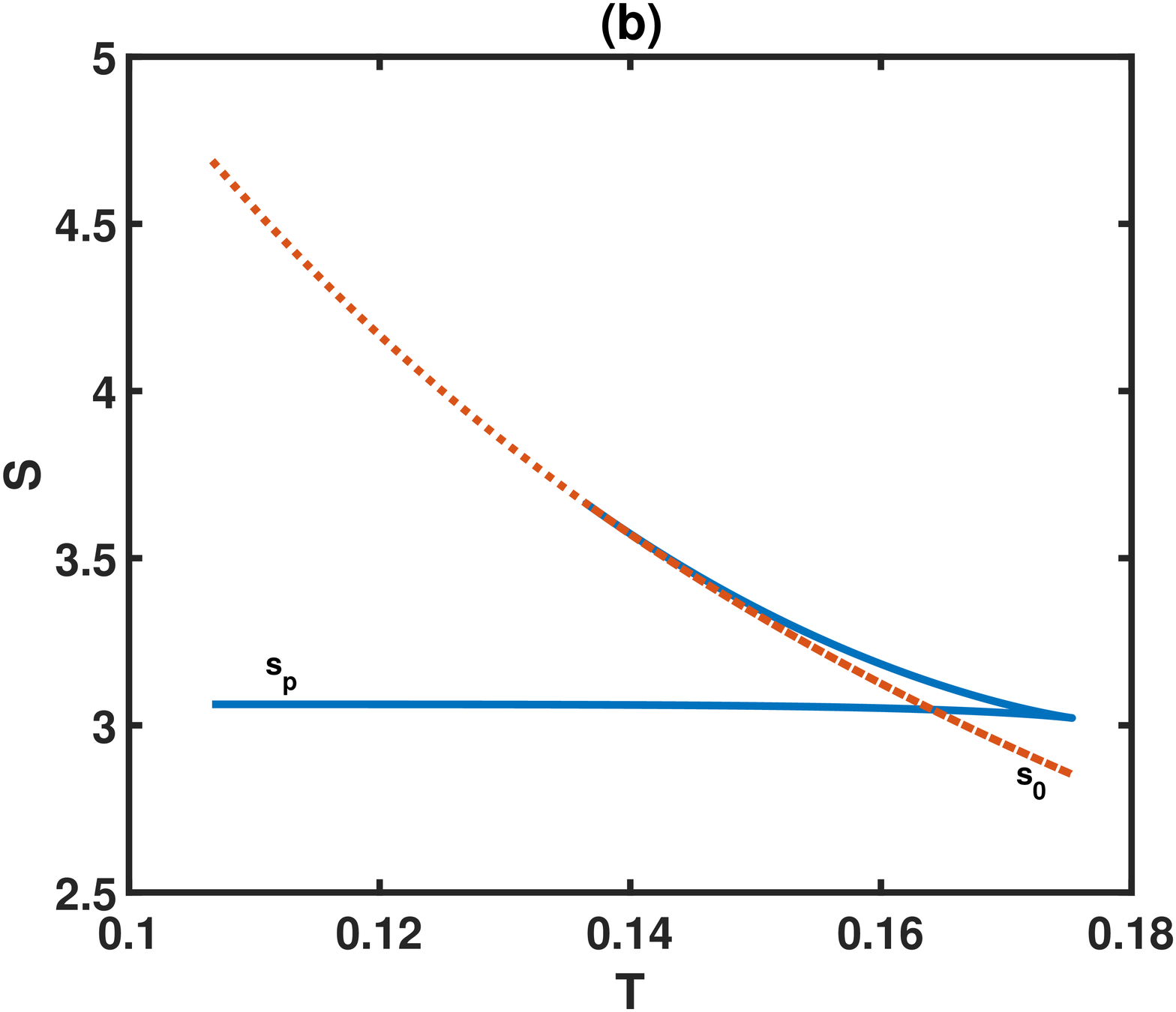}
\caption{(Color online) Plots of the action versus temperature, the dashed line corresponds to the thermodynamic action and the solid line to the periodon action: (a) $\mu = 1$, second-order transition from quantum to thermal regimes, (b) $\mu = 3$, First-order transition from quantum to thermal regimes.  } 
\label{fig3}
\end{figure}

\begin{table}[ht]
\caption{\label{table1} Numerical computation of critical values of $E_1$.} 
\centering 
\renewcommand{\arraystretch}{1.2}
\begin{tabular}{c c c c c c c c c c} 
\hline \hline 
 $\mu^{2}$ & & & Energy $E_{1}$ at  & & & $E_{0}= a_{0}$\\ 
  & & & minimun of $\tau_{p}(E)$  & & & (barrier height)\\  
\hline 
 9 & & & 0.1503 & & & 0.5\\ 
 4 & & & 0.2278 & & & 0.5\\ 
 2 & & & 0.3817 & & & 0.5\\ 
 1.6 & & & 0.4691 & & & 0.5\\ 
 1.501 & & & 0.4994 & & & 0.5\\ [1.ex] 
\hline 
\end{tabular}
\end{table}
The table shows the energy $E_{1}$ approaching the maximum energy $E_{0}$, when $\mu$ tends to $\sqrt{3/2}$. Therefore the critical value of $\mu$ for a transition in quantum tunneling would be $\mu_{c}=\sqrt{3/2}$, and smaller values of $\mu$ should correspond to unphysical values of the energy $E$. Clearly a first-order quantum-classical transition will occur in the parametrized DW model when the deformability parameter lies in the range $\mu > \sqrt{3/2}$. For values of $\mu$ far above the critical threshold $\mu_{c}$ we should have $E_{1}\approx 0$, and $E_{1}\rightarrow E_{0}$ as $\mu$ decreases to $\mu_{c}$. So to say, increasing the deformability parameter above the critical value $\mu_{c}$ should result in a sharper first-order transition \cite{37}.\\  
Fig. \ref{fig2} shows the energy dependence of the period $\tau_{p}$ of the periodon, for two values of the shape deformability parameter $\mu$ selected respectively below and above the critical value $\mu_{c}$. Fig. \ref{fig2}(a) indicates a monotonic decrease of the periodon period with increasing energy for $\mu = 1$, lower than $\mu_{c}$. However, for $\mu = 3$ which is a value greater than $\mu_{c}$, the period has a re-entrant behaviour afer decreasing till a critical value of the energy, as evidenced in fig. \ref{fig2}(b). The re-entrant behaviour of the period actually indicates favorable condition for a first-order transition.
The trend of physical behaviours discussed above and emerging from plots of fig. \ref{fig2}, can also be observed in the corresponding plots of the periodon and the thermodynamic actions shown in fig. \ref{fig3}, still for $\mu = 1$ and $\mu = 3$. 
When the temperature increases one observes a smooth change from $S_{p}$ to $S_{0}$ in fig. \ref{fig3}(a), which is the signature of a second-order transition. In fig. \ref{fig3}(b) the change from the quantum to the classical regime is abrupt, this abrupt change in the temperature dependence of the minimum action fulfills the second criterion proposed by Chudnovsky \cite{8} for a first order-phase transition in temperature. 
\\
For the purpose of mathematical simplicity it can be useful to be able to determine analytically, the critical value $\mu_c$ of $\mu$ for a first-order transition. In this purpose we exploit the proposal of ref. \cite{11}, who extended the criteria for determining the order of transition in quantum tunneling to the dependence of the Euclidean action with the instanton period. Indeed the authors proposed that provided the period $\tau_{p}(E \rightarrow V_{0})$ of the periodon close to the barrier peak can be obtained, the condition $\tau_{p}(E \rightarrow V_{0}) - \tau_{s} < 0$ or $\omega^{2} - \omega_{s}^{2}>0$ will be sufficient for a first order-transition. Here $V_{0}$ and $\tau_{s}$ denote respectively the barrier height and the period of small oscillations around the sphaleron, $\omega$ and $\omega_{s}$ are the corresponding frequencies. In our context the last criterion translates to:
\begin{equation}
V''''(u_{sph},\mu) - \frac{5\left[ V'''(u_{sph},\mu)  \right]^{2}}{3V''(u_{sph},\mu)} < 0,
\label{e18}
\end{equation}
where $u_{sph} = 0$ corresponds to the position of the sphaleron solution. The potential beeing symmetric we have $V'''(0,\mu)=0$, such that the criterion reduces to $V''''(0,\mu)<0$. Using the general expression of the potential given by formula (\ref{e1}), the criterion can be expressed more generally:
\begin{equation}
4\left(\frac{3}{\mu^{2}} - 2 \right)\frac{\alpha(\mu)^{4}a(\mu)}{\mu^{2}}< 0,
\label{e19}
\end{equation}
which suggests a critical value of $\mu_c^2= 3/2$ irrespective the specific forms of $\alpha(\mu)$ and $a(\mu)$. This is in agreement with the results of ref. \cite{26}, where the phase transition in quantum tunneling was investigated with different forms of $a(\mu)$ and $\alpha(\mu)$ corresponding the complete hierarchy of Dikand\'e-Kofan\'e DW potentials. \\

To close this section, we wish to underline that the parametrized DW potential considered in this study is member of the family of potentials changing slowly near the top and the bottom, which were recently designated to stand for ideal candidates for a first order transition \cite{31}. When $\mu\rightarrow 0$, the parametrized DW potential reduces to the $\phi^4$ model that can account only for second-order transitions. As $\mu$ increases, the concave shoulders of the barrier become less pronounced and the narrowest part of the barrier enlarges progressively. So as the shape deformability parameter approaches a critical value $\mu_c$, the barrier top becomes similar to that of a rectanguler barrier and the probability of thermally assisted tunneling, just below the top of the barrier, decreases gradually due to the increase of the tunneling distance. In the region $\mu>\mu_{c}$ this tunneling becomes unfavorable and a first-order transition occurs as a consequence of a direct competition between the groundstate tunneling and thermal activation. The independence of $\mu_{c}$ on the barrier height or the barrier width, highlights the critical role of the shape of the top of the barrier in the transition from quantum to thermal regime.

\section{Statistical mechanics and quasi-exact solvability}
\label{four}
In this section we study the low-temperature statistical mechanics of the parametrized DW potential model (\ref{e1})-(\ref{e2}), paying attention to the canonical partition function and its formulation in terms of an eigenvalue problem within the framework of the  transfer-integral formalism \cite{13a,13,13b,13c,14,15}. To this last point, although the transfer-integral formalism for the family of parametrized DW potentials (\ref{e1}) was discussed in ref. \cite{15}, recent developments on the subject have brought out the possible existence of exact solutions to the transfer-integral operator for parametrized DW potential models. Addressing bistable systems for which the transfer-integral operator corresponding to the $\phi^{4}$ model is known not to have exact solutions, in ref. \cite{32} it was shown that when the $\phi^4$ model is suitably modified to the form  $V(x) = V_{0}[ (n+\xi)x^{2} - (n-\xi)]^{2}$ and provided some conditions on $n$ and $\xi$, the model can be connected to the double sinh-Gordon potential whose transfer-integral operator can be exactly diagonalized for particular values of $n$ and $\xi$. This result make relevant the search for QES potentials in the deformable realm, in the following we address the quasi-exact solvability of the parametric bistable potential in formula (\ref{e1}) with the parameters $\alpha(\mu)$ and $a(\mu)$ given by (\ref{e2}). \\
In the displacive limit, the (dimensionless) Hamiltonian governing the system dynamics in the continuum of space is given by:
\begin{equation}
H = \int dx\left[ \frac{1}{2}\pi^{2} + \frac{1}{2}(\partial_{x}u)^{2}  + V(u,\mu)\right],
\label{e20}
\end{equation}
where $\pi$ is the momentum density conjugate to the displacement field $u$. The solitary-wave solution to the equation of motion i.e.: 
\begin{equation}
\frac{\partial^{2}u}{\partial t^{2}} - \frac{\partial^{2}u}{\partial x^{2}} + V(u,\mu) = 0,
\label{e21}
\end{equation}
derived from the Hamilonian (\ref{e20}) in the displacive regime, is found to be: 
\begin{eqnarray}
u_{c}(s,\mu)&=& \pm  \frac{1}{\alpha(\mu)}\tanh^{-1}\left[\frac{\mu}{\sqrt{1+\mu^{2}}} \tanh \frac{s}{\sqrt{2}d(\mu)}\right], \nonumber \\
\label{e22} \\
   s&=&x-vt, \nonumber
\end{eqnarray}
and corresponds to the vacuum instanton already obtained in formula (\ref{e6}). This solution describes a kink (+) or an antikink (-) with a rest mass:  
\begin{equation}
M_{0}(\mu) = \frac{\sqrt{2a(\mu)}}{2\alpha(\mu)\mu^{2}}\left[ 2\alpha(\mu)(1+\mu^{2}) - sinh(2\mu)\right].
\label{e23}
\end{equation}
For the low-temperature statistical mechanics, we apply the transfer-integral formalism \cite{13,14,15} and using the Hamiltonian (\ref{e20}), we find the following Schr\"odinger-like equation from the transfer-operator eigenvalue equation: 
\begin{equation}
-\frac{1}{2\beta^{2}}\frac{\partial^{2}}{\partial u^{2}}\psi_{j} + a(\mu)\left( \frac{\sinh ^{2}(\alpha(\mu) u)}{\mu^{2}} - 1\right)^{2}\psi_{j} = \epsilon_{j}\psi_{j}.
\label{e24}
\end{equation}
In the thermodynamic limit $N\rightarrow\infty$, only the lowest eigenvalue $\epsilon_{0}$ has a relevant contribution and $Z$ reduces to the classical partition function;
\begin{equation}
Z_c = \left( 2\pi / \beta h\right)^{N}\exp \left( -\beta N \epsilon_{0}\right),
\label{e25}
\end{equation}
where $\beta = (k_{B}T)^{-1}$ and $N$ is the total number of particles in the system. \\
In the most general context, finding exact values for the partition function $Z$ depends on the solvability of the eigenvalue problem (\ref{e24}). To transform the present eigenvalue problem into a form for which the exact solvability is established \cite{39,32}, we use the variable change $z=\alpha(\mu) u$ and introduce the notations $\xi = (1+2\mu^{2})^{-1}$, $E_{j} = 4\mu^{4}\xi^{2}\epsilon_{j}/{a(\mu)}$. With these new variables the eigenvalue problem eq. (\ref{e24}) becomes:
 \begin{equation}
\frac{1}{2\beta^{2}}\frac{\partial^{2}}{\partial z^{2}}\psi_{j} +\frac{a(\mu) }{4\mu^{4}(\alpha(\mu)\xi)^{2}}\left[E_{j} - \left( \xi \cosh(2z) - 1\right)^{2}\right]\psi_{j} = 0.
\label{e26}
\end{equation}
Remark that taking $2\beta = 1$ and rescaling the parameter $E_{j}\rightarrow E_{j}/\eta^{2}$, with $\eta = \sqrt{a(\mu)}/[2\mu^{2}\alpha(\mu)\xi]= n+1$, $n=0,1,2,...$, the eigenvalue problem eq. (\ref{e26}) transforms exactly to the equation treated in ref. \cite{39}. Instructively, in this last work it was possible to obtain exact expressions for some eigenfunctions and their corresponding eigenenergies for some energy levels $n$. Following his scheme in our context amounts to taking one energy level at a fixed temperature, which will yield few values of $\mu$ for which the system is QES. However what we really want is to obtain eigenstates at different temperatures, for each value of $\mu$. This is possible by considering the specific case where the temperature is related to the shape deformability parameter through:
 \begin{equation}
\beta^{2} = \frac{2\mu^{4}(\alpha(\mu)\xi)^{2}}{a(\mu)}q^{2},
\label{e27}
\end{equation} 
where $q$ is a positive integer we shall call the "temperature order". Substituting (\ref{e27}) into (\ref{e26}) we obtain:
 \begin{equation}
\frac{\partial^{2}}{\partial z^{2}}\psi_{j} + q^{2} \left[E_{j} - \left( \xi \cosh(2z) - 1\right)^{2}\right]\psi_{j} = 0.
\label{e28}
\end{equation}
Equation (\ref{e28}) dsecribes a QES system analog to that studied in refs. \cite{39,32}, for the energy level $n=1$. We are interested in solutions that vanish in the limit $z\rightarrow \pm \infty$, therefore we can readily define:
\begin{equation}
\psi(z) = r(z)\exp\left[ -z_{0}\cosh(2z)\right],
\label{e29}
\end{equation}
where the function $r(z)$ is a polynomial expressed as a linear combinnation of $\cosh m_{p}z$ or $\sinh  m_{p}z$ and their powers, with $z_{0}$ and $ m_{p}$ ( $p=0,1,2,...$) constant quantities to be determined. \\
The exact expressions for the (unnormalized) groundstate wavefunctions, and of the associated eigenenergies, are found for the first four values of $q$ and are:

\begin{enumerate}
 \item[] $q=1$:
 \begin{eqnarray}
 \psi_{0}(u) &=& \exp\left[ -\frac{\cosh(2\alpha(\mu)u)}{2(1+2\mu^{2})}\right], \nonumber \\
 \epsilon_{0} &=& \frac{a(\mu)}{4\mu^{4}}\left[ 1 + (1+2\mu^{2})^{2}\right]. \label{e30}
 \end{eqnarray}
 
 \item[] $q=2$:
  \begin{eqnarray}
 \psi_{0}(u) &=& \cosh(\alpha(\mu)u)\exp\left[ -\frac{\cosh(2\alpha(\mu)u)}{1+2\mu^{2}}\right], \nonumber \\
 \epsilon_{0} &=& \frac{a(\mu)}{16\mu^{4}}\left[ 3(1+2\mu^{2})^{2}-8\mu^{2}\right]. \label{e31}
 \end{eqnarray}
 
  \item[] $q=3$:
 \begin{eqnarray}
 \psi_{0}(u)&=& \lbrack\frac{6}{1+2\mu^{2}}+ \left( 1 + \sqrt{1+\frac{36}{(1+2\mu^{2})^{2}}}\right)\nonumber \\
 &\times&\cosh(2\alpha(\mu)u)\rbrack\exp\left[-\frac{3\cosh(2\alpha(\mu)u)}{2(1+2\mu^{2})}\right], \nonumber \\
 \epsilon_0 &=& \frac{a(\mu)}{36\mu^{4}}\lbrack 9 -2(1+2\mu^{2})\sqrt{(1+2\mu^{2})^{2} +36} \nonumber \\
 &+& 7(1+2\mu^{2})^{2}\rbrack. \label{e33}
 \end{eqnarray}
 
  \item[] $q=4$:
  \begin{eqnarray}
 \psi_{0}(u)&=& 2\lbrack\frac{6\cosh(\alpha(\mu)u)}{1+2\mu^{2}} \nonumber \\
 &+& \left(2\mu^{2}-1 +\sqrt{12-8\mu^2+(1+2\mu^{2})^2}\right) \nonumber \\
 &\times& \frac{\cosh(3\alpha(\mu)u)}{1+2\mu^{2}}\rbrack \exp\left[-\frac{2\cosh(2\alpha(\mu)u)}{1+2\mu^{2}}\right], \nonumber \\
 \epsilon_{0} &=& \frac{a(\mu)}{16\mu^{4}}\lbrack 2 -4\mu^{2}-(1+2\mu^{2}) \nonumber \\
 &\times& \sqrt{(1+2\mu^{2})^{2} -8\mu^{2} +12} +\frac{11}{4}(1+2\mu^{2})^{2}\rbrack. \nonumber \\ \label{e35}
 \end{eqnarray}
 \end{enumerate}
 
\begin{figure} 
\includegraphics[width=3.5in,height=2.5in]{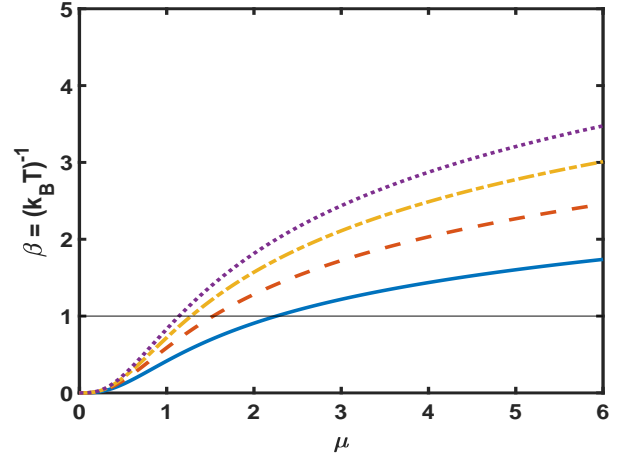}
\caption{(Color online) Variation of the inverse temperature $\beta$ as a function of the shape deformability parameter $\mu$, for four different values of $q$ according to the quasi-exact solvability condition (\ref{e27}): $q=1$ (Solid line), $q=2$ (Dashed line), $q=3$ (Dash-dotted line), $q=4$ (Dotted line). The horizontal line stands for the energy barrier taken to be $a_{0}=1$.} 
\label{fig4}
\end{figure}

\begin{figure} 
\includegraphics[width=3.5in,height=2.5in]{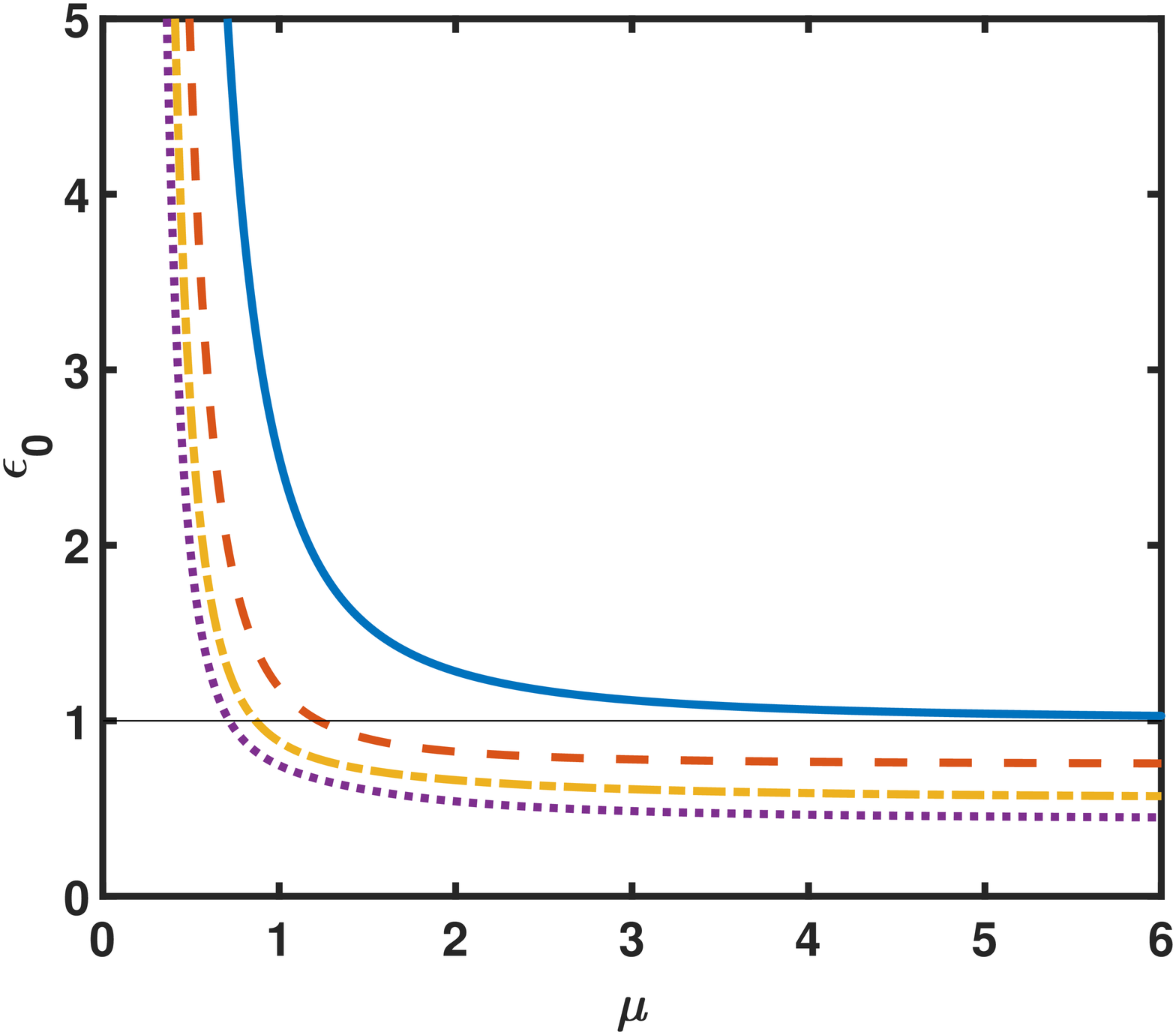}
\caption{(Color online) Variation of the groundstate energy $\epsilon_0$ with the shape deformability parameter $\mu$, for four different values of $q$ (corresponding to four different temperatures) namely: $q=1$ (Solid line), $q=2$ (Dashed line), $q=3$ (Dash-dotted line), $q=4$ (Dotted line). The horizontal line stands for the energy barrier $a_0$ here fixed as $a_0=1$.} 
\label{fig5}
\end{figure}

\begin{figure} 
\includegraphics[width=3.5in,height=2.in]{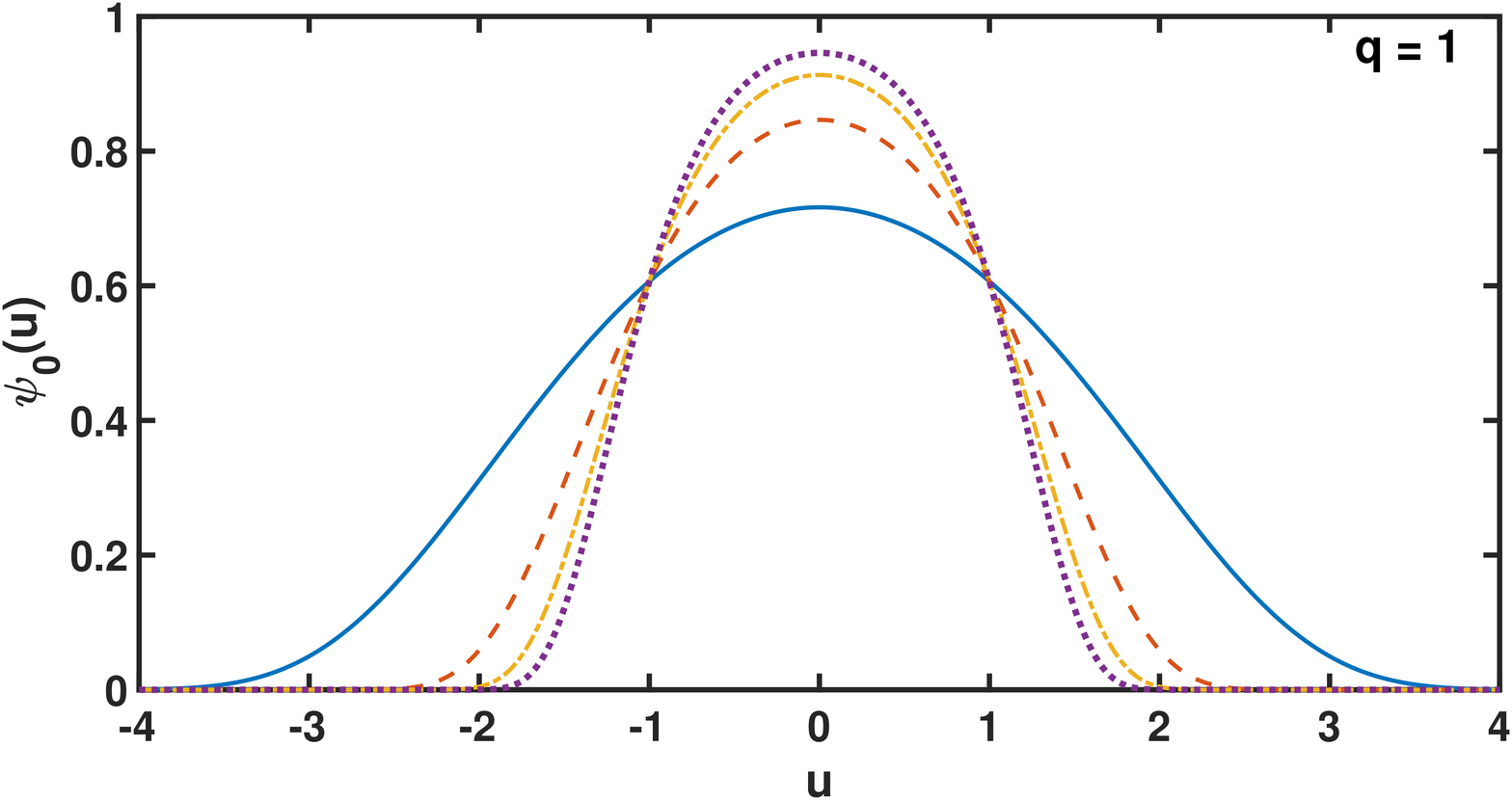}
\includegraphics[width=3.5in,height=2.in]{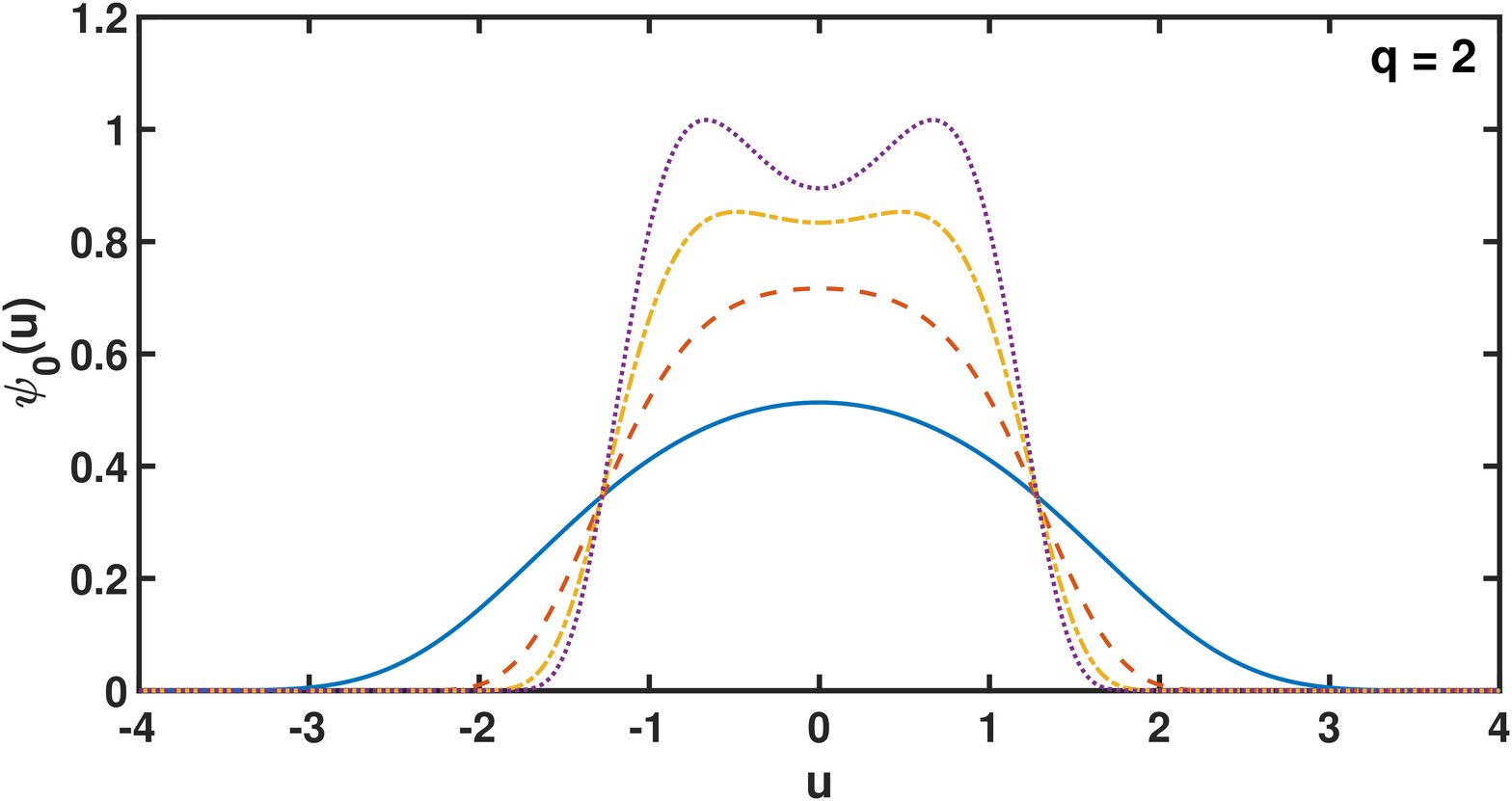}
\includegraphics[width=3.5in,height=2.in]{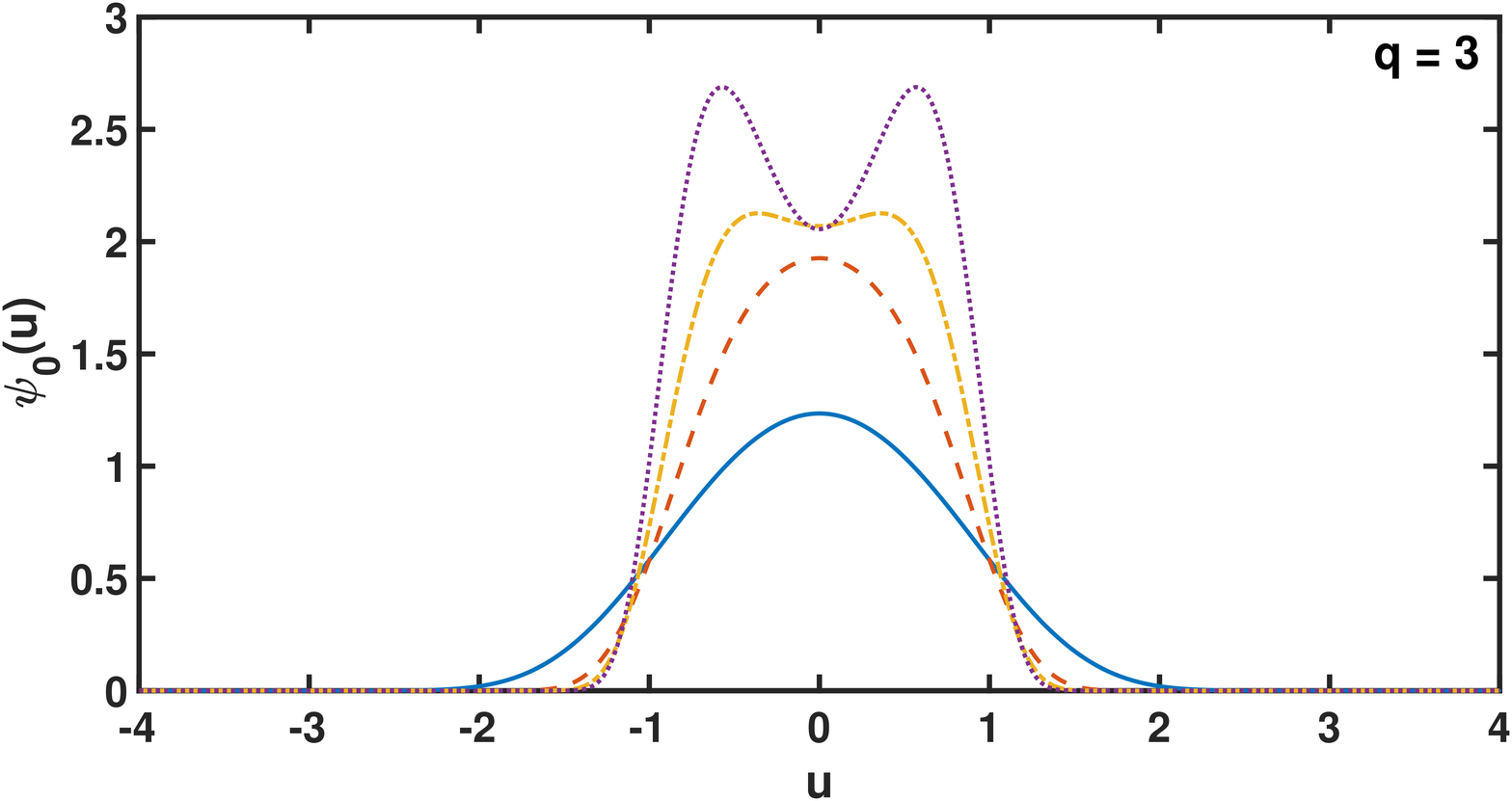}
\includegraphics[width=3.5in,height=2.in]{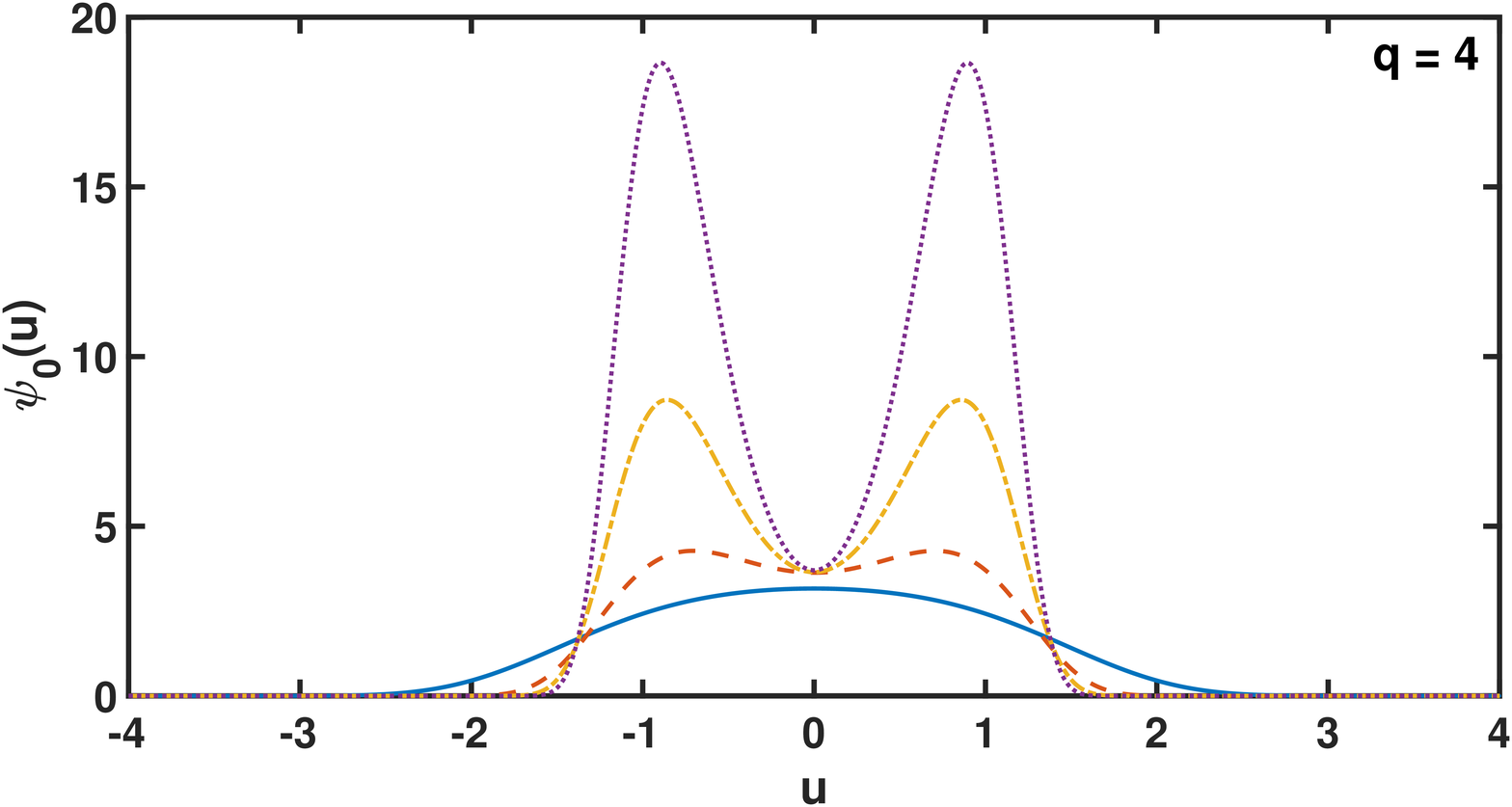}
\caption{(Color online) Groundstate wavefunctions in position space at four different temperatures, for $\mu = 0.5$ (Solid line), $\mu = 1$ (Dashed line), $\mu = 1.5$ (Dot-dashed line),$\mu = 2$ (Dashed line).} 
\label{fig6}
\end{figure}

Eigenfunctions and eigenvalues for some higher energy levels are given in the appendix. \\
The condition for quasi-exact solvability of the system at several temperatures is defined by relation (\ref{e28}), this relation sets the dependence of the temperature (in unit of the Boltzmann constant) on the shape deformability parameter $\mu$ and is illustrated in Fig. \ref{fig4}. It is remarkable on the figure that in the limit $\mu \rightarrow 0$, the quasi-exact integrability condition requires $T\rightarrow \infty$. This is in excellent agreement with the transfer-operator problem for the $\phi^{4}$ model  not having exact solutions at finite temperatures. 
We also note on fig. \ref{fig4}  that for small values of $\mu$, the four temperatures are far greater than the energy barrier $a_0$ that in the unit of Boltzmann constant unit has the sense of the transition temperature. As $\mu$ gradually increases, the temperatures obtained from the largest to the lowest $q$ steadily decrease and respectively go far below the energy barrier. For sufficiently larga values of $\mu$, the exact solvability condition thus holds at four temepratures below the symmetry breaking temperature. \\
The free energy is strongly related to the groundstate energy, the influence of $\mu$ on the groundstate energies and on their relative positions with respect to the energy barrier, is represented in fig. \ref{fig5}.
The groundstate energies are infinitely large for $\mu \rightarrow 0$, and decrease with an increase of $\mu$ irrespective of $q$. However, combining the influence of $\mu$ together with the choice of $q$ affects the position of the energy level with respect to the energy barrier. Taking $q=1$ for illustration, the groundstate energy decreases drastically but will always remain above the energy barrier. This drastic decrease is also observed for higher values of $q$, but when $\mu$ rises beyond a specific value, notably $\mu_{s}= \sqrt{3/2}, \sqrt{3/4}$ and $\mu_{s}\simeq 0.717$ for $q=2$, $q=3$ and $q=4$ respectively, the grounstate drops below the energy barrier and tends to a finite value namely  $\epsilon_{0}(\mu\rightarrow \infty) =(3/4)a_0$ for $q=2$, $\epsilon_{0}(\mu\rightarrow \infty) =(5/9)a_{0}$ for $q=3$ and $\epsilon_{0}(\mu\rightarrow \infty)= (7/16)a_{0}$ for $q=4$. Moreover, for a fixed value of $\mu$ lower energy levels are obtained by increasing $q$.\\
Also of relevant interest to the statistical mechanics of the system, as for instance in computating correlation functions and  correlation lengths, is the fact that with the knowledge of exact groundstate wavefunctions and the corresponding eigen energies at several temperatures for arbitrary values of the deformability parameter $\mu$, quantities such as the probability density become easier to formulate analytically. Indeed the probability density associated with the classical field $u$ is nothing else but the square of the normalized groundstate wavefunction, reason why the influence of shape deformability on the probability density will be qualitatively the same as the influence of this parameter on the groundstate wavefunction. The groundstate wavefunctions (un normalized) for four different temeperatures are plotted in fig. \ref{fig6}, considering different values of the shape deformability parameter $\mu$. 
To understand different profiles of the probability density emerging from fig. \ref{fig6}, it is useful to combine features related to the deformability of the parametriwed DW potential and general properties of probability densities for bistable systems. To this last point, in fig. \ref{fig5} we have seen that for small values of $\mu$ the groundstate energies were higher than the energy barrier $a_0$ and hence correspond to a high-temperature regime. So in this regime the Schr\"odinger pseudoparticles have sufficient energy to cross the energy barrier and move freely from one well to another, behaving as if trapped instead by a single-well potential. The full state space is then covered with power law and the whole space is probable, with a maximum probability density at the barrier peak. As $\mu$ increases, the probability density at the four temperatures experiences increase in amplitude and a decrease in width, while still showing a single peak. This is actually justified by the fact that an increase of $\mu$ enhances the potential confinement by strengthening the steepness of the reflective walls, thus restricting  the attainable space to the region covered by just the valleys and the energy barrier. The groundstate energy beeing high the energy barrier remains the most probable state. For $q = 1$ the groundstate energy lies above the energy barrier, independently of $\mu$. Therefore, even a lowering of temeprature by increase in $\mu$ will still yield a probability density with a single peak, located at the barrier (top graph in fig. \ref{fig6}). However, for choices of $q$ greater than one this behaviour is observed only for $\mu\leq\mu_{S}$. When $\mu$ rises in the range $\mu>\mu_{S}$, the groundstate energy falls below the energy barrier $a_0$ and a decrease in temperature below the transition temperature restricts the transitions form one well to another. The pseudoparticles are consequently more confined in the potential wells, such that the valleys become more probable. In fig. \ref{fig6} this is illustrated by the probability density showing two peaks, each located in the neighborhood of the degenerate minima of the parametrized DW potential. 
\begin{figure} 
\includegraphics[width=3.5in,height=2.5in]{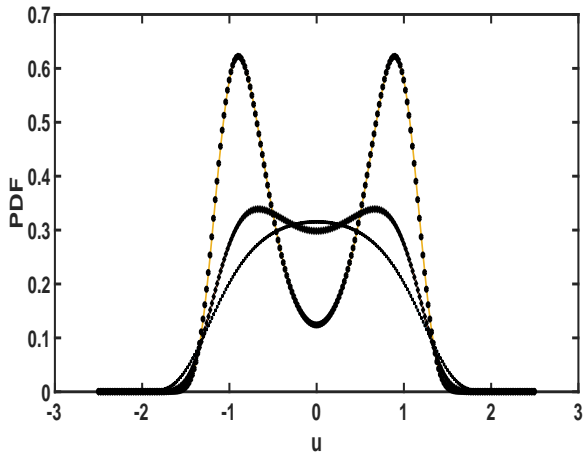}
\caption{(Color online) Comparision of the exact (Solid Line) probability density (PDF for probability density function) and the results from Langevin simulations, for $\mu=2$ and for values of $\beta$ obtained for $q=1$ (stars), $q=2$ (diamond), and $q=4$ (dot). Note the excellent agreement.} 
\label{fig7}
\end{figure}
To enrich the understanding of the temperature dependence of the probability density for a fixed $\mu$ but at different temperatures, in fig. \ref{fig7} we represent the groundstate probability density for $\mu=2$ and for three distinct temperatures at which the QES condition holds. We note the continuous shift from a single peak behavior to a two-peak behavior with decreasing temperature. Remark that the change in the peak width with temperature is different from that of the $\phi^{4}$ model. Instructively the exact probability densities here are superimposed with that obtained from direct simulations of the following additive-noise Langevin equation for the model:
 \begin{equation}
\partial^{2}_{tt}u - \partial^{2}_{xx}u +\eta\partial_{t}u + dV(u,\mu)/du = F(x,t),
 \label{e36}
 \end{equation}
 where the Gaussian white noise $F$ and the viscosity $/eta$ are related by the fluctuation-dissipation theorem:
 \begin{equation}
\langle F(x,t)F(x',t')\rangle = 2\eta\beta^{-1}\delta (x-x')\delta (t-t').
 \label{e37}
 \end{equation}
Using standard techniques \cite{40} to solve the lattice versions of the above continuous equation, the results were sampled in time to obtain time averaged probability densities after random initial conditions had been driven to equilibrium. In the simulations we employed both Euler and Runge-Kutta schemes, with a lattice size of typically 350K points taking a lattice spacing of $k= 0.02$ and a time step $h = 0.001$. As seen in fig. \ref{fig7}, the exact probability densities and the results from simulations show excellent agreement. This agreement between the analytical and numerical results suggests that the groundstate energies can be numerically computed using the corresponding probability densities at any temperature where QES condition holds, for a wide range of shape deformability parameter thereby obtaining thermodynamic quantities such as the internal energy, the free energy and the entropy. 

\section{Summary and conclusion} 
\label{five}
We have investigated the influence of shape deformability of a parametrized DW potential, on the transition from quantum tunneling to thermally induced escape and on the statistical mechanics of bistable systems with deformable DW substrate potentials. Systems with this feature abound in nature, ranging from biology to soft matters such as polymers, molecular crystals and hydrogen-bonded ferroelectrics and antiferroelectrics. In these last systems in particular, isotope effects cause changes in bond lengths and characteristic parameters (the barrier height, minima positions, etc.) of the one-body DW potential governing symmetry breaking induced by proton displacements. By using different methods, we have shown that there exists a critical value of the shape deformability parameter $\mu$ aobve which a first-order transition in quantum tunneling can occur in the system. \\
 The possibility to study the statistical mechanics of the system via the transfer-integral operator method was addressed, with a highlight on condition for quasi-exact solvability of the associate pseudoparticle's Schr\"odinger equation. We obtained groundstate wavefunctions and the corresponding eigen energies for temperatures determined by the quasi-exact solvability condition, and discussed their dependences on the shape deformability parameter. We obtained that for each value of $\mu$ the quasi-exact solvability of the model allows exact computation of the eigenstates at various temperatured. These temperatures are obtined above and below the transition temperature, depending on values of the shape deformability parameter. The influence of shape deformability of the DW potential on the groundstate wavefunctions has been examined. Our analytical results show a striking agreement with large scale Langevin simulations, implying that the study of probability density-based thermodynamics via Langevin simulations turns out to be quite manageable for bistable systems with the current parametrized DW potential.
 
\appendix
\section*{Appendix}
Exact solutions to the eigenvalue problem eq. (\ref{e28}):
\begin{itemize}
\item[]  $q=1$:
 \begin{eqnarray}
 \psi_{0}(\phi) &=& \exp\left[ -\frac{\cosh(2\alpha(\mu)\phi)}{2(1+2\mu^{2})}\right], \nonumber \\
 \epsilon_{0} &=& \frac{a(\mu)}{4\mu^{4}}\left[ 1 + (1+2\mu^{2})^{2}\right]
 \label{e38}
 \end{eqnarray}
\item[] $q=2$:
 \begin{eqnarray}
 \psi_{0}(\phi) &=& \cosh(\alpha(\mu)\phi)\exp\left[ -\frac{\cosh(2\alpha(\mu)\phi)}{1+2\mu^{2}}\right], \nonumber \\
 \epsilon_{0} &=& \frac{a(\mu)}{16\mu^{4}}\left[ 3(1+2\mu^{2})^{2}-8\mu^{2}\right]. 
 \label{e39}
 \end{eqnarray}
\begin{eqnarray}
\psi_{1}(\phi) &=& \sinh(\alpha(\mu)\phi)\exp\left[ -\frac{\cosh(2\alpha(\mu)\phi)}{1+2\mu^{2}}\right], \nonumber \\
\epsilon_{1} &=& \frac{a(\mu)}{16\mu^{4}}\left[ 3(1+2\mu^{2})^{2}+8(1+\mu^{2})\right],
\label{e40}
 \end{eqnarray}
with  $\epsilon_{1} - \epsilon_{0}= \frac{a(\mu)}{4\mu^{4}} (1+2\mu^{2})$.
\item[] $q=3$:
\begin{eqnarray}
 \psi_{0}(\phi) &=& \lbrack \frac{6}{1+2\mu^{2}}+ \left( 1 + \sqrt{1+\frac{36}{(1+2\mu^{2})^{2}}}\right)\nonumber \\
 &\times& \cosh(2\alpha(\mu)\phi)\rbrack \exp\left[ -\frac{3\cosh(2\alpha(\mu)\phi)}{2(1+2\mu^{2})}\right], \nonumber \\
 \epsilon_{0} &=& \frac{a(\mu)}{36\mu^{4}}\lbrack 9 -2(1+2\mu^{2})\sqrt{(1+2\mu^{2})^{2} +36}\nonumber \\
 &+& 7(1+2\mu^{2})^{2}\rbrack.
 \label{e42}
 \end{eqnarray}
 \begin{eqnarray}
 \psi_{1}(\phi) &=& \sinh(2\alpha(\mu)\phi)\exp\left[ -\frac{3\cosh(2\alpha(\mu)\phi)}{2(1+2\mu^{2})}\right], \nonumber \\
 \epsilon_1 &=& \frac{a(\mu)}{36\mu^{4}}\left[ 9 + 5(1+2\mu^{2})^{2}\right]. \label{e43}
 \end{eqnarray}
 \begin{eqnarray}
 \psi_{2}(\phi) &=& \lbrack \frac{6}{1+2\mu^{2}}- \left( \sqrt{1+\frac{36}{(1+2\mu^{2})^{2}}}-1\right) \nonumber \\
&\times& \cosh(2\alpha(\mu)\phi)\rbrack \exp\left[ -\frac{3\cosh(2\alpha(\mu)\phi)}{2(1+2\mu^{2})}\right], \nonumber \\
 \epsilon_{2} &=& \frac{a(\mu)}{36\mu^{4}}\lbrack 9 +2(1+2\mu^{2})\sqrt{(1+2\mu^{2})^{2} +36} \nonumber \\
 &+& 7(1+2\mu^{2})^{2}\rbrack,
 \label{e45}
 \end{eqnarray}
with:
\begin{equation}
\epsilon_{1} - \epsilon_{0}= \frac{a(\mu)}{18\mu^{4}}\left[ (1+2\mu^{2})\sqrt{(1+2\mu^{2})^{2}+36} - 2(1+2\mu^{2})^{2}\right], \label{ma1}
\end{equation}
and: 
\begin{equation}
\epsilon_{2} - \epsilon_{0}=  \frac{a(\mu)}{9\mu^{4}}\left[ (1+2\mu^{2})\sqrt{(1+2\mu^{2})^{2}+36}\right]. \label{ma}
\end{equation}

\item[] $q=4$
 \begin{eqnarray}
 \psi_{0}(\phi) &=& 2\lbrack \frac{ 6\cosh(\alpha(\mu)\phi)}{1+2\mu^{2}} \nonumber \\ 
 &+& \left(2\mu^{2}-1 +\sqrt{12-8\mu^{2}+(1+2\mu^{2})^{2}}\right) \nonumber \\
 &\times& \frac{\cosh(3\alpha(\mu)\phi)}{1+2\mu^{2}}\rbrack \exp\left[ -\frac{2\cosh(2\alpha(\mu)\phi)}{1+2\mu^{2}}\right], \nonumber \\
 \epsilon_{0} &=& \frac{a(\mu)}{16\mu^{4}}\lbrack 2 -4\mu^{2}-(1+2\mu^{2})\nonumber \\
 &\times& \sqrt{(1+2\mu^{2})^{2} -8\mu^{2} +12}+\frac{11}{4}(1+2\mu^{2})^{2}\rbrack. \nonumber \\ 
 \label{e47}
 \end{eqnarray}
 \begin{eqnarray}
 \psi_{1}(\phi) &=& 2\lbrack \frac{6\sinh(\alpha(\mu)\phi)}{1+2\mu^{2}} \nonumber \\
 &+& \left(2\mu^{2}+3 +\sqrt{20+8\mu^{2}+(1+2\mu^{2})^{2}}\right)\nonumber \\
 &\times& \frac{\sinh(3\alpha(\mu)\phi)}{1+2\mu^{2}}\rbrack \nonumber \\
  &\times& \exp\left[ -\frac{2\cosh(2\alpha(\mu)\phi)}{1+2\mu^{2}}\right], \nonumber \\
 \epsilon_{1} &=& \frac{a(\mu)}{16\mu^{4}}\lbrack 6 +4\mu^{2}-(1+2\mu^{2})\nonumber \\
 &\times& \sqrt{(1+2\mu^{2})^{2} +8\mu^{2} +20}+\frac{11}{4}(1+2\mu^{2})^{2}\rbrack. \nonumber \\
 \label{e49}
 \end{eqnarray}
 \begin{eqnarray}
 \psi_2(\phi) &=& 2\lbrack \frac{6\cosh(\alpha(\mu)\phi)}{1+2\mu^{2}} \nonumber \\
 &+& \left(2\mu^{2}-1 -\sqrt{12-8\mu^{2}+(1+2\mu^{2})^{2}}\right)\nonumber \\
 &\times& \frac{\cosh(3\alpha(\mu)\phi)}{1+2\mu^{2}}\rbrack \nonumber \\
  &\times& \exp\left[ -\frac{2\cosh(2\alpha(\mu)\phi)}{1+2\mu^{2}}\right], \nonumber \\
 \epsilon_2 &=& \frac{a(\mu)}{16\mu^{4}}\lbrack 2 -4\mu^{2}+(1+2\mu^{2})\nonumber \\
 &\times& \sqrt{(1+2\mu^{2})^{2} -8\mu^{2} +12}+\frac{11}{4}(1+2\mu^{2})^{2}\rbrack, \nonumber \\ 
 \label{e51}
 \end{eqnarray}
 with: 
 \begin{eqnarray}
\epsilon_{1} - \epsilon_{0} &=& \frac{a(\mu)}{16\mu^{4}}\lbrack 4+ 8\mu^{2} \nonumber \\
&+& (1+2\mu^{2})\sqrt{(1+2\mu^{2})^{2} -8\mu^{2} +12}\rbrack \nonumber \\ 
&-& \frac{a(\mu)}{16\mu^{4}}\left[(1+2\mu^{2})\sqrt{(1+2\mu^{2})^{2} +8\mu^{2} +20}\right], \nonumber \\
\label{eqad}
 \end{eqnarray} 
and:
\begin{equation}
\epsilon_{2} - \epsilon_{0} = \frac{a(\mu)}{8\mu^{4}}\left[ (1+2\mu^{2})\sqrt{(1+2\mu^{2})^{2} -8\mu^{2} +12} \right]. \label{eqasd}
\end{equation}
\end{itemize}

\section*{Acknowledgments}
A. M. Dikand\'e is grateful to the Alexander von Humboldt foundation for sponsoring his visit at the Max-Planck Institute for the Physics of Complex Ssystems (MPIPKS) in Dresden, Germany where part of this work was done. He thanks Pr. Holger Kantz for accepting to be his host, and Pr. Peter Fulde for his constant encouragements.

\section*{Author contribution statement}
F. N. N. did analytical calculations and carried out numerical simulations under supervision of A. M. D. and C. T., A. M. D. developed some of the codes used in simulations and wrote the manuscript, C. T. contributed to discussion of results and
writing of the manuscript.

\end{document}